\documentclass[twocolumn,preprintnumbers,amsmath,amssymb]{revtex4}
\usepackage{graphicx}  % needed for figures
\usepackage{dcolumn}   % needed for some tables
\usepackage{bm}        % for math
\usepackage{amssymb}   % for math
\begin{document}

\title{The symbiotic contact process: phase transitions, hysteresis cycles, and bistability}

\author{C. I. N. Sampaio Filho$^1 \footnote{Correspondence to: cesar@fisica.ufc.br}$, T. B. dos Santos$^{1}$, N. A. M. Ara\'{u}jo$^{1,2}$, H. A. Carmona$^1$, A. A. Moreira$^1$, J. S. Andrade Jr.$^{1}$}

\affiliation{$^1$Departamento de F\'{i}sica, Universidade Federal do Cear\'a, 
  60451-970 Fortaleza, Cear\'a, Brasil\\
  $^2$ Departamento  de F\'{i}sica, Faculdade de Ci\^{e}ncias, Universidade de Lisboa, P-1749-016 Lisboa, Portugal, and
  Centro de F\'{i}sica Te\'{o}rica e Computacional, Universidade de Lisboa, P-1749-016 Lisboa, Portugal}

\begin{abstract}
We performed Monte Carlo simulations of the symbiotic contact process on different spatial dimensions ($d$). On the complete and random graphs (infinite dimension), we observe hysteresis cycles and bistable regions, what is consistent with the discontinuous absorbing-state phase transition predicted by mean-field theory. By contrast, on a regular square lattice, we find no signs of bistability or hysteretic behavior. This result suggests that the transition in two dimensions is rather continuous. Based on our numerical observations, we conjecture that the nature of the transition changes at the upper critical dimension ($d_c$), from continuous ($d<d_c$) to discontinuous ($d>d_c$).
\end{abstract}

\maketitle

\section{\label{sec:level1}Introduction} 

Improving our understanding of absorbing-state phase transitions in non-equilibrium systems is of great importance, not only because they occur in a variety of problems, but also display critical behavior and universality~\cite{Odor2004,Dickman2005,Henkel2009}.
Absorbing states are those at which the dynamics is suppressed and no further changes occur. Examples of these states were found in models of epidemic spreading, opinion formation~\citep{Anteneodo2017}, population dynamics~\cite{Sarkar2015}, diffusion-limited aggregation~\cite{Kartha2016,Iannini2017}, traffic~\cite{Antonov2016}, and other non-equilibrium systems~\cite{Gutierrez2017,Takeuchi2007}.
Most of these models are characterized by a continuous phase transition that falls into the Directed Percolation (DP) universality class~ \cite{Grassberger1982,Janssen1981,Odor2004}. However, absorbing phase transitions might also be discontinuous. Examples include, the single-species restrictive contact process models, such as the quadratic contact process (QCP)~\cite{Guo2007,Liu2007, Guo2009,Silva2012,Varghese2013,Fiore2014, Fiore2014a, Pastor-Satorras2015}, the Ziff-Gulari-Barshad (ZGB) model for catalysis~\cite{Ziff1986,fiorePRE2015,Oliveira2016a}, and ballistic deposition with anisotropic interactions~\cite{diasPRE2014,araujo2015kinetic}.

The two-species contact process 2SCP was introduced by Oliveira \textit{et al.}~\cite{Oliveira2012} to study the effects of symbiotic interactions in the contact process (CP)~\cite{Harris1974}. As in CP, in 2SCP the dynamics of each species evolves through sequences of creation and annihilation, but the rate of annihilation is reduced in the presence of a second species. Oliveira~\textit{et al.} have shown that, in the mean-field limit, the absorbing-state phase transition in 2SCP becomes discontinuous for a wide range of the symbiotic interaction strengths~\cite{Oliveira2012}. However, no evidence of a discontinuous transition in two dimensions has been observed from numerical simulations on a square lattice~\cite{Oliveira2012,Oliveira2014}. Here, we combine Monte Carlo simulations and a mean-field calculation to study the nature of the referred transition. We focus on the stability of the steady state and hysteretic behavior. In the mean-field limit, we confirm that the absorbing-state phase transition might be discontinuous, while in two dimensions it is always continuous and belongs to the Directed Percolation universality class~\cite{Luebeck2003a,Henkel2009}.

The paper is organized as follows. In Section II we describe the 2SCP model and derive the phase diagram and bistable regions in the mean-field regime. The simulation results for different underlying networks are presented in Section III. In Section IV we draw some final conclusions.

\section{\label{sec:level2}The Two-species contact process} 

In the 2SCP two species ($A$ and $B$) are considered. Each site of a network is either empty or occupied by only one $A$-particle, only one $B$-particle, or two different particles. At a given instant $t$, the state of the site $i$ is characterized by a pair of variables $\left( \sigma_{i}(t), \eta_{i}(t) \right)$, where $\sigma_{i}(t) = 1$ ($\eta_i(t)=1$) if the site is occupied by one $A$-particle ($B$-particle) or $\sigma_{i}(t) = 0$ ($\eta_i(t)=0$) otherwise. The transition from $(0,\eta_{i}) \to (1,\eta_{i})$ occurs at rate $\lambda r_{A}$, being $r_{A}$ the fraction of nearest neighbors (NN) occupied by $A$-particles, independently of $\eta_{i}$. In the same way, the transition from $(\sigma_{i},0) \to (\sigma_{i},1)$ occurs at rate $\lambda r_{B}$, with $r_{B}$ the fraction of NN occupied by $B$-particles, independently of $\sigma_{i}$.  The annihilations $(1,0) \to (0,0)$ and $(0,1) \to (0,0)$ occur at rate of unity, while the ones $(1,1) \to (0,1)$ and $(1,1) \to (1,0)$ occur at rate $ \mu \leq 1$, i.e., the rate of annihilation is reduced on sites occupied by particles of both species (symbiosis).

In the 2SCP the symbiotic interaction favors the persistence of the doubled occupied sites, and the critical reproduction rate $\lambda_{c}$ decreases as the parameter $\mu$ is reduced. Moreover, a continuous phase transition in the Directed Percolation universality class, is observed  for $\mu > 1/2$. The upper critical dimension $d_{u}$ of this model is the same of the ordinary CP, namely, $d_{u} = 4$.  From the mean-field equations, it was previously found~Ref.~\cite{Oliveira2012} that the phase transition is discontinuous for $\mu < 1/2$, with $\mu = 1/2$ identified as the tricritical point. 
In what follows, we study the stability of the steady state when the transition is discontinuous.

The state where $(\sigma_i,\eta_i)=\left(0,0\right)$ for all $i$ is absorbing. At $\lambda_{c}(\mu)$ the system undergoes an absorbing phase transition~\citep{Oliveira2012,Oliveira2014}. The mean-field theory for the 2SCP was first derived in Ref.~\cite{Oliveira2012}, assuming spatial homogeneity. Defining $p_0$, $p_A$, $p_B$, and $p_{AB}$ as probabilities for a given site to be empty, occupied by only one $A$-particle, only one $B$-particle, or by both species, respectively, they studied the effect of symbiotic interactions by seeking a symmetric solution $p_{A} = p_{B} = p$, which obeys, 
\begin{equation}
 \frac{dp}{dt}= \lambda(1 - p_{AB} - 3p)(p + p_{AB}) + \mu{p_{AB}} - p,
 \label{eq01}
\end{equation}
and
\begin{equation}
 \frac{dp_{AB}}{dt}= 2\lambda{p}(p + p_{AB}) - 2\mu p_{AB},
 \label{eq02}
\end{equation} 
using the constrain $p_{0} = 1 - 2p - p_{AB}$. The absorbing state corresponds to $p=0$ and $p_{AB} = 0$. The active stationary solutions $\left(dp/dt = 0  \mbox{ and } dp_{AB}/dt = 0 \right)$ are given by   

\begin{equation}
p^{\pm} =  \frac{\mu}{2\lambda (1-\mu)} \left[ 2(1 - \mu) - \lambda \pm \sqrt{\lambda^{2} - 4\mu(1 - \mu)} \right],
\label{eq03}
\end{equation}
and
\begin{equation}
 p^{\pm}_{AB} = \frac{\lambda (p^{\pm})^{2}}{\mu - \lambda p^{\pm}}.
\label{eq04}
\end{equation}
We define the order parameter as the density of particles $\rho$, which depends on both parameters $\left(p^{0,\pm},p_{AB}^{0,\pm}\right)$. Therefore, taking into account the steady-state solutions, we calculate $\rho$ in the mean-field limit, for all values of the parameters $\lambda$ and $\mu$. We focus in the limit $\mu < 1/2$, where the 2SCP undergoes a discontinuous phase transition~\cite{Oliveira2012}. Since only $\rho \ge 0$ has physical meaning, there are three solutions, namely, 

\begin{equation}
\rho_{\mbox{absorbing}} = 2p^{0} + 2p_{AB}^{0},
\label{eq05}
\end{equation}
\begin{equation}
\rho_{\mbox{active}} = 2p^{+} + 2p_{AB}^{+}, 
\label{eq06}
\end{equation}
and
\begin{equation}
\rho_{\mbox{unstable}} = 2p^{-} + 2p_{AB}^{-},  
\label{eq07}
\end{equation}
\begin{figure}[t]
\includegraphics*[width=\columnwidth]{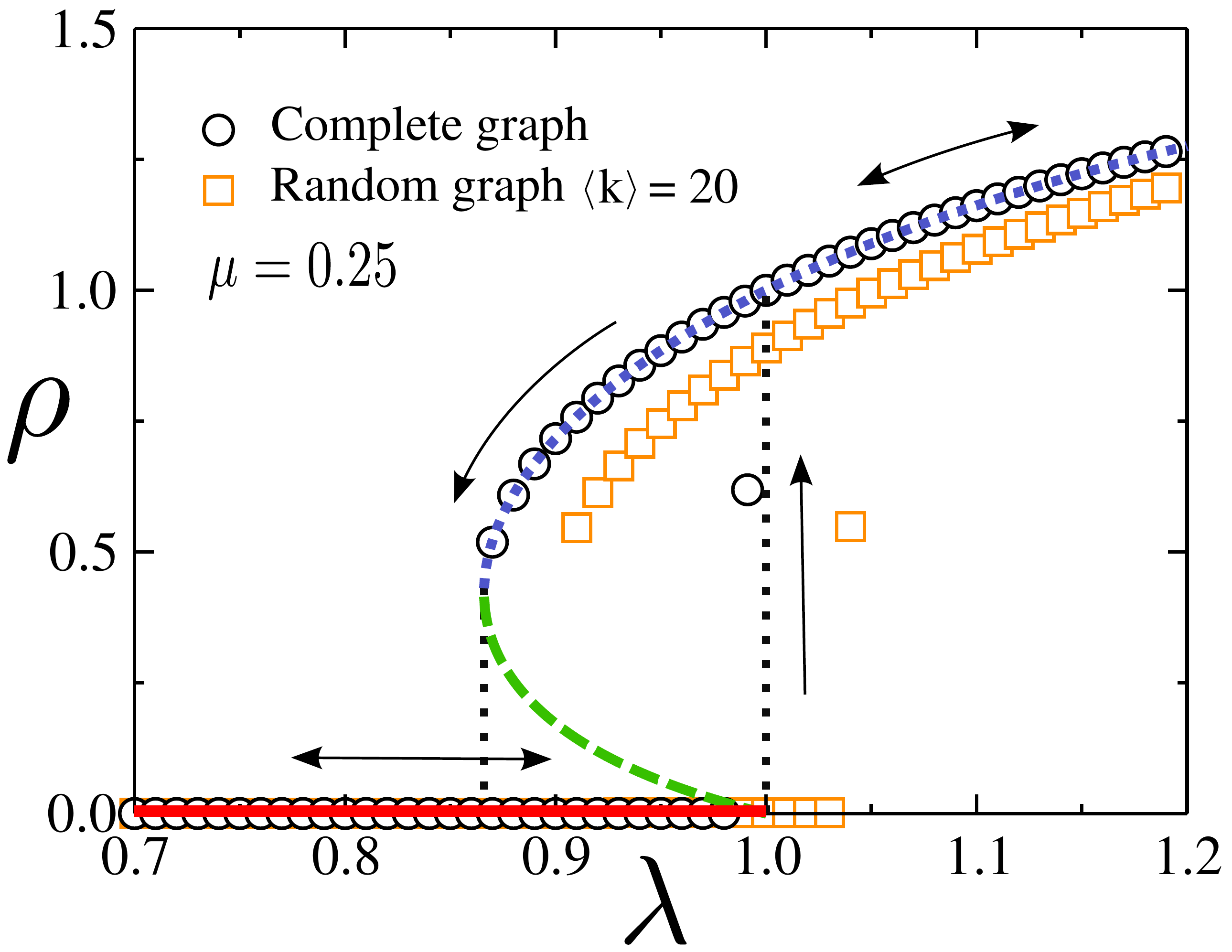}
\caption{Hysteretic cycle for $\mu = 1/4$, obtained from the mean-field calculation. Symbols represent simulations performed in complete graphs (circles) and random graphs (rectangles). The highlighted arrows indicate the direction of the cycle. The solution $\rho = 0$ (the red continuous line) corresponds to the stable absorbing state. The solution $\rho = 2p^{-} + 2p_{AB}^{-}$ (the dashed green line) is unstable for any value of $\lambda$. Finally, $\rho  =  2p^{+} + 2p_{AB}^{+}$ represents the stable active solution (the dotted blue line).}
\label{fig01}
\end{figure}
\begin{figure}[htb]
\includegraphics*[width=\columnwidth]{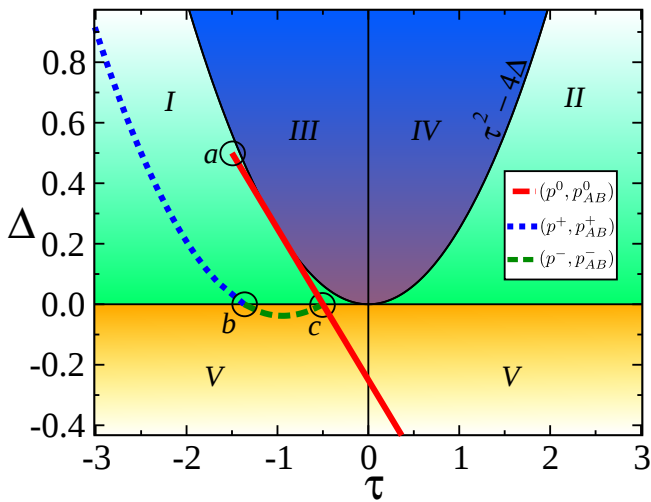}
\caption{(Color online) The $(\tau,\Delta)$ stability diagram of the mean-field solutions for $\mu = 1/4$. The regions $I,II,III,IV,$ and $V$ correspond to, respectively, the stable nodes, unstable nodes, stable spirals, unstable spirals, and saddle points. The solution $\rho = 0$ (continuous red line) is stable if $0 \leq\lambda \leq 1$, since, from this condition, the absorbing solution lies in the region $I$. For $\lambda > 1$ the absorbing solution lies in the region $V$, being therefore unstable. The active solution (dotted blue line) $\rho = 2p^{+} + 2p_{AB}^{+}$, lies in the region $I$ if $\lambda \geq \lambda_{c}\left(\mu=1/4\right) = \sqrt{3/4}$. The solution $\rho = 2p^{-} + 2p_{AB}^{-}$ (dashed green line) is unconditionally unstable, since for any value of $\lambda$ this solution lies in the region $V$. }
\label{fig02}
\end{figure} 
\noindent where the indexes ``absorbing'', ``active'', and ``unstable'' refer to the type of solution, as discussed below.

One signature of a discontinuous transition is the presence of hysteretic behavior. Figure~\ref{fig01} shows the hysteresis cycle obtained from the mean-field calculation for the case $\mu = 1/4$. The solution $\rho = 0$ (continuous red line) corresponds to the absorbing phase. The solutions $\rho =  2p^{+} + 2p_{AB}^{+}$ (dotted blue line) and $\rho =  2p^{-} + 2p_{AB}^{-}$ (dashed green line) are physical if $\lambda \geq \lambda_{c}(\mu)$ and $\lambda_{c}(\mu) <\lambda < 1$, respectively, with $\lambda_{c} = 2\sqrt{\mu\left(1-\mu\right)}$. Otherwise, $\rho$ would admit complex values. For the case $\mu = 1/4$, we have $\lambda_{c} = \sqrt{3/4}$. However, as discussed next, the solution given by Eq.~(\ref{eq07}) is always unstable, while the stability of other solutions depends on the values of $\lambda$ and $\mu$.

To analyze the stability of each solution, we consider the Jacobian matrix. The system described by Eqs.~(\ref{eq01}) and (\ref{eq02}) can be written as $\frac{dp}{dt} = f(p,p_{AB})$ and $\frac{dp_{AB}}{dt} = g(p,p_{AB})$. The Jacobian matrix is then
\begin{equation}
A(p,p_{AB}) = 
\left( \begin{array}{cc}
\frac{\partial{f}}{\partial{p}} & \frac{\partial{f}}{\partial{p_{AB}}}  \\
\frac{\partial{g}}{\partial{p}} & \frac{\partial{g}}{\partial{p_{AB}}}  \\
\end{array} \right).
\end{equation}
The trace $\tau(\lambda,\mu)$ and the determinant $\Delta(\lambda,\mu)$ of the matrix $A$ for each steady-state solution are
\begin{equation}
 \tau(p^0,p_{AB}^0) = \lambda - 1 - 2\mu,
 \label{eq08}
\end{equation}
\begin{equation}
 \Delta(p^0,p_{AB}^0) = 2\mu(1-\lambda),
 \label{eq09}
\end{equation}
\begin{equation}
 \tau(p^{\pm},p_{AB}^{\pm}) = 2\mu - \lambda - 1 \pm2\sqrt{4\mu^2 - 4\mu + \lambda^2},
 \label{eq10}
\end{equation}
and
\begin{equation}
 \Delta(p^{\pm},p_{AB}^{\pm}) = 4\mu^2 - 4\mu + \lambda^2 \pm(2\mu - \lambda)\sqrt{4\mu^2 - 4\mu + \lambda^2}.
 \label{eq11}
\end{equation}	
This analysis can be summarized in Fig.~\ref{fig02}, which shows the $(\tau,\Delta)$ stability diagram of the Jacobian matrix for $\mu = 1/4$ (the same parameters as in Fig.~\ref{fig01}). The diagram is divided into five regions. The regions $I$ and $II$ correspond to the stable and unstable nodes, respectively. The regions $III$ and $IV$ correspond to the stable and unstable spirals. Finally, the region $V$ corresponds to the saddle points, namely, an unstable region. The solution $\rho = 0$ (dashed red line) is conditionally stable, since for $0 \leq \lambda \leq 1$, this solution belongs to the region $I$ of stable nodes. However, for $\lambda > 1$, the absorbing solution is a saddle node (region $V$) and becomes unstable. The solution $\rho = 2p^{+} + 2p_{AB}^{+}$, corresponding to an active phase, is stable if $\lambda \geq \lambda_{c}=\sqrt{3/4}$. Notice that in the range $\sqrt{3/4} < \lambda < 1$ either absorbing or actives phases are stable. This range, therefore, bounds the bistable region. Finally, the solution $\rho = 2p^{-} + 2p_{AB}^{-}$ is unconditionally unstable, since for any value of $\lambda$ this solution lies in the region $V$ of saddle nodes.

\section{Complete and random graphs}

In order to check the histeretic behavior predicted by the mean-field calculation, we performed Monte Carlo simulations of the symbiotic contact process on complete and random graphs. We considered the algorithm described in Ref.~\cite{Oliveira2012}. Accordingly, we define $\delta t$ as the time increment associated to a given step in the 2SCP simulation and $N_{s}$ and $N_{d}$ as the number of sites occupied by one or two species, respectively. At each time step, we choose one of the following events: 

\begin{itemize}
\item creation attempt at a site occupied only by a single species, with probability $\lambda N_{s}\delta t$;
\item creation attempt at a site occupied by both species, with probability $2\lambda N_{d}\delta t$;
\item annihilation of a particle at a site occupied only by a single species, with probability $N_{s}\delta t$;
\item annihilation of a particle at a site occupied by both species, with probability $2\mu N_{d}\delta t$.
\end{itemize}
Since the probabilities are normalized, $1/\delta t  = \lambda N_{p} + N_{s} + 2\mu N_{d}$, where $N_{p} = N_{s} + 2N_{d}$ is the total number of particles. Moreover, we take $\delta t = 1/N_{p}$ on the graphs of $N_{p}$ active nodes, such that a Monte Carlo step corresponds to one attempt event per node, on average~\cite{Dickman1998}. Using this algorithm, we follow the time evolution of the 2SCP. However, due to finite-size effects, the absorbing configuration can always be reached, even for $\lambda > \lambda_{c}(\mu)$, what would immediately suppress the dynamics~\cite{Dickman1998,Sander2016}. To circumvent this problem, every time that an absorbing configurations is generated, we perform a spontaneous creation of two particles, one of each species, in sites chosen at random. Notice that this method guarantees that there will be at least one particle of each species at all times.

\begin{figure}[t]
\includegraphics*[width=\columnwidth]{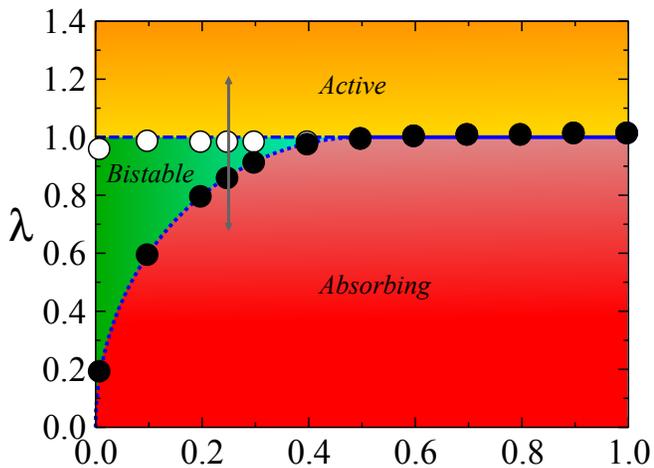}
\caption{(Color online) The phase diagram of the symbiotic contact process. Three phases can be identified, Active, Bistable, and Absorbing. The continuous, dashed and dotted lines represent solutions of the mean-field equations. For $\mu > 1/2$, the system undergoes a continuous phase transition between the active and absorbing phase. For $\mu < 1/2$, the system describes a discontinuous phase transition, defining a bistable region where the active and absorbing phases are both stables. The highlighted arrow indicates the direction of the transition for the case $\mu = 0.25$, where the bistable region is identified in the hysteretic cycle shown in Fig.~\ref{fig02}. The open and solid symbols represent the critical reproduction rate $\lambda_c$ obtained from simulations performed in complete graphs, with $N = 5\times 10^{4}$. For the open symbols, the initial configuration is in the absorbing state, while  for the solid symbols the fully occupied system defines the initial configuration.}
\label{fig03}
\end{figure}

A complete graph is defined as a structure where each node interacts with all others. Figure~\ref{fig01} shows the density of particles for a complete graph (circles) and random graphs (rectangles) of $N=5\times 10^4$ nodes. The hysteresis cycle was obtained for a fixed value of $\mu=1/4$. For each value of $\lambda$, we allowed the dynamics to evolve for $\Delta t$ MCS. Next, we increased and decreased $\lambda$ by constant intervals $\Delta \lambda$, and simulated the dynamics starting from the previous configuration, for each value of $\lambda$~\cite{vladimirPRE2009}. Each data point is an average over $10^{2}$ independent configurations. As can be seen in the Fig.~\ref{fig01}, the results for the complete graph are in good agreement with the mean-field solutions. Moreover, both in the complete and random graphs the nature of the hysteretic behavior is consistent with a discontinuous transition. 

Figure~\ref{fig03} shows the phase diagram of the 2SCP obtained for a complete graph and mean-field solutions, where active, bistable, and absorbing phases are identified. For $\mu>1/2$, the 2SCP  undergoes a continuous absorbing-state phase transition. The solid symbols represent the critical parameter $\lambda_{c}$ obtained by the ratio cumulant~\cite{Dickman2005}. The continuous line represents the respective mean-field solution. For $\mu <1/2$, the system undergoes a discontinuous phase transition with a bistable phase, where both the active and absorbing phases are stable. The initial configuration here is an absorbing state for the open symbols and a fully occupied state for the solid symbols. Notice the agreement between the simulated data (symbols) and the mean-field solutions.

\section{Regular square lattice}

\begin{figure}[t]
\includegraphics*[width=\columnwidth]{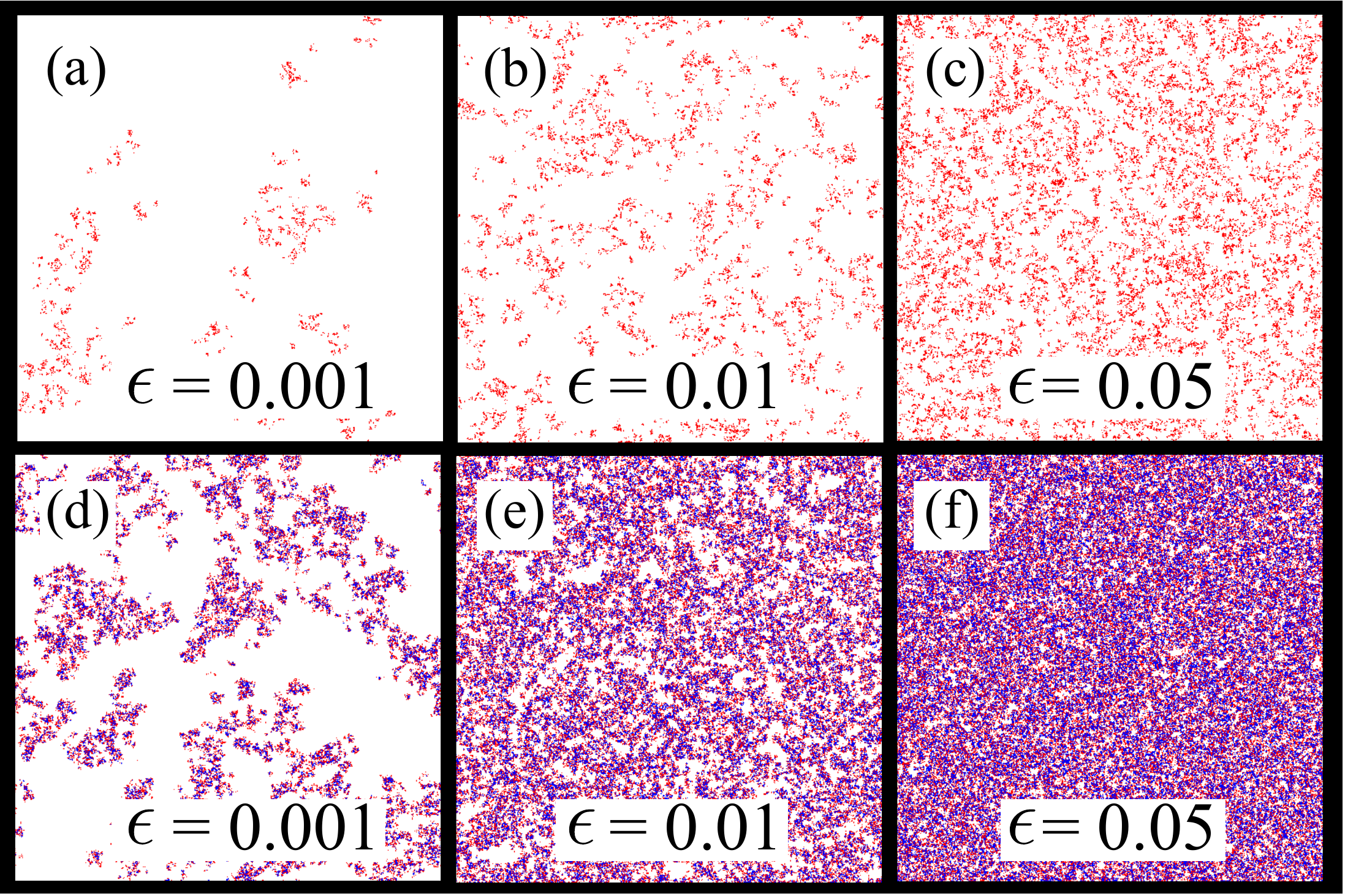}
\caption{Snapshots of the ordinary contact process (from a to c) and the symbiotic contact process (from d to f) on a square regular lattice  at the steady state, shifted by the quantities $\epsilon = 0.001, 0.01, 0.05$ from the respective critical creation rate $\lambda_{c}$.}
\label{fig04}
\end{figure}
  
We now consider the 2SCP on regular square lattice. Figure \ref{fig04} shows snapshots of the ordinary and symbiotic contact processes at the steady state for $\epsilon = 0.001, 0.01, 0.05$, where $\epsilon = \lambda - \lambda_{c}$ and  $\lambda_{c} = 1.6488(1)$ \cite{Dickman1998} for the ordinary contact process (Figs.~\ref{fig04}a-c) and $\lambda_{c}(\mu = 0.25) = 1.13730(5)$ \cite{Oliveira2012} for the 2SCP (Figs.~\ref{fig04}d-f). For both models, the same method described for the complete and random graphs was used to avoid the absorbing state. Notice that, for any value of $\epsilon$, the density for actives sites of the 2SCP is always greater than that for the ordinary contact process.

\begin{figure}[b]
\includegraphics*[width=\columnwidth]{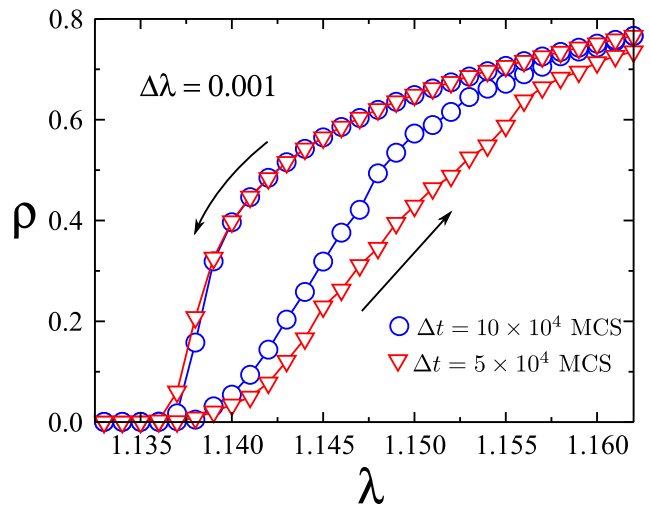}
\caption{Hysteresis cycles of the order parameter $(\rho)$ in terms of the creation rate $\lambda$, for $\mu = 0.25$, on a square lattice of linear size $L = 200$. Here we consider $\Delta t = 5 \times 10^{4}$ MCS (red triangles) and $\Delta t = 10\times 10^{4}$ MCS (blue circles) as time increments.  For each cycle, the control parameter $\lambda$ is increased and decreased in the range $1.0 \leq \lambda \leq 1.20$ at the constant intervals $\Delta \lambda = 0.001$.}
\label{fig05}
\end{figure}

\begin{figure*}[t]
\includegraphics*[width=17cm,height=5cm]{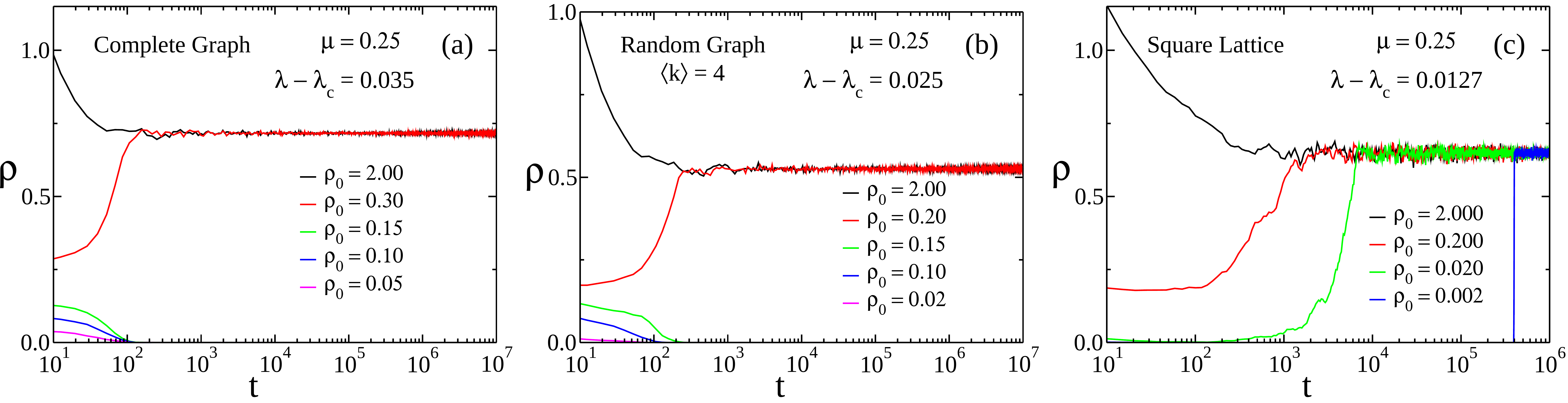}
\caption{The dependence of the initial density $\rho_{0}$ of particles  on stationary state. Accordingly, we have fixed the values of $\lambda$ and $\mu$ to identify a possible bistable behavior. a) Complete graph, $\mu = 0.25$ and $\lambda - \lambda_c =  0.035$. b) Random graph,  $\mu = 0.25$ and $\lambda - \lambda_c =  0.025$. c) Regular square lattice, $\mu = 0.25$ and $\lambda - \lambda_c =  0.0127$.}
\label{fig06}
\end{figure*} 

To determine the order of the phase transition on regular lattices, we analyze the hysteresis cycles. We employ the same algorithms used in Section III to produce the $QS$ states and the hysteresis cycles. Figure~\ref{fig05} shows, for $\mu = 0.25$, the order parameter for two cycles in the creation rate $\lambda$. For each cycle, the control parameter $\lambda$ is varied in the range $1.0 \leq \lambda \leq 1.20$ at constant intervals $\Delta \lambda = 0.001$. Note that, as we double the value of $\Delta t$, the width of the cycle is decreased. This indicates an absence of hysteretic behavior for $\Delta t \to \infty$. Moreover, since 
the time necessary to reach the steady state diverges at the critical region in the thermodynamic limit, a system that undergoes a continuous phase transition to an absorbing phase should exhibit an hysteresis cycle when the control parameter is varied around its critical value \cite{Takeuchi2007,vladimirPRE2009}.

The absence of bistability for the 2SCP on regular square lattices can be studied by evaluating the role of initial conditions on the stationary state, as described in Ref.~\cite{vladimirPRE2009}. Considering different values of the initial density $\rho_{0}$ of particles, with fixed values of $\lambda$ and $\mu$, we can evaluate the stability of each state. Figure~\ref{fig06} shows the results obtained on a regular square lattice, and on complete and random graphs. The values of $\lambda$ and $\mu$ are in a range where a possible bistable region is identified. As expected, for the complete and random graphs (Figs.~\ref{fig06}a and b) the stable phase depends on the initial condition considered, reflecting the presence of a bistable region between the absorbing and active phases. However, on a square lattice (Fig.~\ref{fig06}c) the active phase is always stable, for all considered initial conditions. This indicates that bistability is not observed for 2SCP on two-dimensional lattices. Moreover, we conjecture that the 2SCP always have a continuous phase transition below the upper critical dimension.

\section{Conclusions}

We have revisited the symbiotic contact process, where two species interact via a reduced death rate $\mu$, that describes the dynamics of doubled occupied sites, but individually, the dynamics of each species is described by an ordinary contact process. We have shown that, by using a suitable method to generate the quasistationary state (QS), the simulations performed on complete graphs are in accordance with the mean-field solutions. Precisely, these solutions reveal a discontinuous phase transitions, with hysteretic behavior and a bistable phase, where the absorbing and the actives phases are both stables. A bistable region also is detected on random graphs. Considering simulations on regular square lattices, we show the absence of hysteretic behavior and bistable regions, being these properties consistent with a continuous phase transition. Moreover, we conjecture that the 2SCP always undergoes a continuous phase transition 
for any spatial dimension below the upper critical dimension, but above one-dimensional systems.

\begin{acknowledgments}
We thank the Brazilian agencies CNPq, CAPES, FUNCAP, and the National Institute of Science and Technology for Complex Systems for financial support. NAMA acknowledges financial support from the Portuguese Foundation for Science and Technology (FCT) under Contract no. UID/FIS 00618/2013. 
\end{acknowledgments}


\begin{thebibliography}{32}
\expandafter\ifx\csname natexlab\endcsname\relax\def\natexlab#1{#1}\fi
\expandafter\ifx\csname bibnamefont\endcsname\relax
  \def\bibnamefont#1{#1}\fi
\expandafter\ifx\csname bibfnamefont\endcsname\relax
  \def\bibfnamefont#1{#1}\fi
\expandafter\ifx\csname citenamefont\endcsname\relax
  \def\citenamefont#1{#1}\fi
\expandafter\ifx\csname url\endcsname\relax
  \def\url#1{\texttt{#1}}\fi
\expandafter\ifx\csname urlprefix\endcsname\relax\def\urlprefix{URL }\fi
\providecommand{\bibinfo}[2]{#2}
\providecommand{\eprint}[2][]{\url{#2}}

\bibitem[{\citenamefont{\'{O}dor}(2004)}]{Odor2004}
\bibinfo{author}{\bibfnamefont{G.}~\bibnamefont{\'{O}dor}},
  \bibinfo{journal}{Rev. Mod. Phys.} \textbf{\bibinfo{volume}{76}},
  \bibinfo{pages}{663} (\bibinfo{year}{2004}).

\bibitem[{\citenamefont{Dickman and Marro}(2005)}]{Dickman2005}
\bibinfo{author}{\bibfnamefont{R.}~\bibnamefont{Dickman}} \bibnamefont{and}
  \bibinfo{author}{\bibfnamefont{J.}~\bibnamefont{Marro}},
  \emph{\bibinfo{title}{Nonequilibrium Phase Transitions in Lattice Models}}
  (\bibinfo{publisher}{Cambridge University Press}, \bibinfo{year}{2005}), ISBN
  \bibinfo{isbn}{978-0521019460}.

\bibitem[{\citenamefont{Henkel et~al.}(2009)\citenamefont{Henkel, Hinrichsen,
  and L\"ubeck}}]{Henkel2009}
\bibinfo{author}{\bibfnamefont{M.}~\bibnamefont{Henkel}},
  \bibinfo{author}{\bibfnamefont{H.}~\bibnamefont{Hinrichsen}},
  \bibnamefont{and} \bibinfo{author}{\bibfnamefont{S.}~\bibnamefont{L\"ubeck}},
  \emph{\bibinfo{title}{Non-Equilibrium Phase Transitions. Volume 1}}
  (\bibinfo{publisher}{Springer-Verlag GmbH}, \bibinfo{year}{2009}), ISBN
  \bibinfo{isbn}{978-1-4020-8764-6}.

\bibitem[{\citenamefont{Anteneodo and Crokidakis}(2017)}]{Anteneodo2017}
\bibinfo{author}{\bibfnamefont{C.}~\bibnamefont{Anteneodo}} \bibnamefont{and}
  \bibinfo{author}{\bibfnamefont{N.}~\bibnamefont{Crokidakis}},
  \bibinfo{journal}{\textit{Phys. Rev. E}} \textbf{\bibinfo{volume}{95}},
  \bibinfo{pages}{042308} (\bibinfo{year}{2017}).

\bibitem[{\citenamefont{Sarkar}(2015)}]{Sarkar2015}
\bibinfo{author}{\bibfnamefont{N.}~\bibnamefont{Sarkar}},
  \bibinfo{journal}{\textit{Phys. Rev. E}} \textbf{\bibinfo{volume}{92}},
  \bibinfo{pages}{042110} (\bibinfo{year}{2015}).

\bibitem[{\citenamefont{Kartha and Banpurkar}(2016)}]{Kartha2016}
\bibinfo{author}{\bibfnamefont{M.~J.} \bibnamefont{Kartha}} \bibnamefont{and}
  \bibinfo{author}{\bibfnamefont{A.~G.} \bibnamefont{Banpurkar}},
  \bibinfo{journal}{\textit{Phys. Rev. E}} \textbf{\bibinfo{volume}{94}},
  \bibinfo{pages}{062108} (\bibinfo{year}{2016}).

\bibitem[{\citenamefont{Iannini and Dickman}(2017)}]{Iannini2017}
\bibinfo{author}{\bibfnamefont{M.~L.~L.} \bibnamefont{Iannini}}
  \bibnamefont{and} \bibinfo{author}{\bibfnamefont{R.}~\bibnamefont{Dickman}},
  \bibinfo{journal}{Phys. Rev. E} \textbf{\bibinfo{volume}{95}},
  \bibinfo{pages}{022106} (\bibinfo{year}{2017}).

\bibitem[{\citenamefont{Antonov et~al.}(2016)\citenamefont{Antonov,
  Hnati{\v{c}}, Kapustin, Lu{\v{c}}ivjansk{\'y}, and
  Mi{\v{z}}i{\v{s}}in}}]{Antonov2016}
\bibinfo{author}{\bibfnamefont{N.~V.} \bibnamefont{Antonov}},
  \bibinfo{author}{\bibfnamefont{M.}~\bibnamefont{Hnati{\v{c}}}},
  \bibinfo{author}{\bibfnamefont{A.~S.} \bibnamefont{Kapustin}},
  \bibinfo{author}{\bibfnamefont{T.}~\bibnamefont{Lu{\v{c}}ivjansk{\'y}}},
  \bibnamefont{and}
  \bibinfo{author}{\bibfnamefont{L.}~\bibnamefont{Mi{\v{z}}i{\v{s}}in}},
  \bibinfo{journal}{\textit{Phys. Rev. E}} \textbf{\bibinfo{volume}{93}},
  \bibinfo{pages}{012151} (\bibinfo{year}{2016}).

\bibitem[{\citenamefont{Guti\'{e}rrez et~al.}(2017)\citenamefont{Guti\'{e}rrez,
  Simonelli, Archimi, Castellucci, Arimondo, Ciampini, Marcuzzi, Lesanovsky,
  and Morsch}}]{Gutierrez2017}
\bibinfo{author}{\bibfnamefont{R.}~\bibnamefont{Guti\'{e}rrez}},
  \bibinfo{author}{\bibfnamefont{C.}~\bibnamefont{Simonelli}},
  \bibinfo{author}{\bibfnamefont{M.}~\bibnamefont{Archimi}},
  \bibinfo{author}{\bibfnamefont{F.}~\bibnamefont{Castellucci}},
  \bibinfo{author}{\bibfnamefont{E.}~\bibnamefont{Arimondo}},
  \bibinfo{author}{\bibfnamefont{D.}~\bibnamefont{Ciampini}},
  \bibinfo{author}{\bibfnamefont{M.}~\bibnamefont{Marcuzzi}},
  \bibinfo{author}{\bibfnamefont{I.}~\bibnamefont{Lesanovsky}},
  \bibnamefont{and} \bibinfo{author}{\bibfnamefont{O.}~\bibnamefont{Morsch}},
  \bibinfo{journal}{\textit{Phys. Rev. A}} \textbf{\bibinfo{volume}{96}},
  \bibinfo{pages}{041602} (\bibinfo{year}{2017}).

\bibitem[{\citenamefont{Takeuchi et~al.}(2007)\citenamefont{Takeuchi, Kuroda,
  Chat\'{e}, and Sano}}]{Takeuchi2007}
\bibinfo{author}{\bibfnamefont{K.~A.} \bibnamefont{Takeuchi}},
  \bibinfo{author}{\bibfnamefont{M.}~\bibnamefont{Kuroda}},
  \bibinfo{author}{\bibfnamefont{H.}~\bibnamefont{Chat\'{e}}}, \bibnamefont{and}
  \bibinfo{author}{\bibfnamefont{M.}~\bibnamefont{Sano}},
  \bibinfo{journal}{\textit{Phys. Rev. Lett.}} \textbf{\bibinfo{volume}{99}},
  \bibinfo{pages}{234503} (\bibinfo{year}{2007}).

\bibitem[{\citenamefont{Grassberger}(1982)}]{Grassberger1982}
\bibinfo{author}{\bibfnamefont{P.}~\bibnamefont{Grassberger}},
  \bibinfo{journal}{Z. Phys. B Cond. Matt.} \textbf{\bibinfo{volume}{47}},
  \bibinfo{pages}{365} (\bibinfo{year}{1982}).

\bibitem[{\citenamefont{Janssen}(1981)}]{Janssen1981}
\bibinfo{author}{\bibfnamefont{H.~K.} \bibnamefont{Janssen}},
  \bibinfo{journal}{Z. Phys. B Cond. Matt.} \textbf{\bibinfo{volume}{42}},
  \bibinfo{pages}{151} (\bibinfo{year}{1981}).

\bibitem[{\citenamefont{Guo et~al.}(2007)\citenamefont{Guo, Liu, and
  Evans}}]{Guo2007}
\bibinfo{author}{\bibfnamefont{X.}~\bibnamefont{Guo}},
  \bibinfo{author}{\bibfnamefont{D.-J.} \bibnamefont{Liu}}, \bibnamefont{and}
  \bibinfo{author}{\bibfnamefont{J.~W.} \bibnamefont{Evans}},
  \bibinfo{journal}{\textit{Phys. Rev. E}} \textbf{\bibinfo{volume}{75}},
  \bibinfo{pages}{061129} (\bibinfo{year}{2007}).

\bibitem[{\citenamefont{Liu et~al.}(2007)\citenamefont{Liu, Guo, and
  Evans}}]{Liu2007}
\bibinfo{author}{\bibfnamefont{D.-J.} \bibnamefont{Liu}},
  \bibinfo{author}{\bibfnamefont{X.}~\bibnamefont{Guo}}, \bibnamefont{and}
  \bibinfo{author}{\bibfnamefont{J.~W.} \bibnamefont{Evans}},
  \bibinfo{journal}{\textit{Phys. Rev. Lett.}} \textbf{\bibinfo{volume}{98}},
  \bibinfo{pages}{050601} (\bibinfo{year}{2007}).

\bibitem[{\citenamefont{Guo et~al.}(2009)\citenamefont{Guo, Liu, and
  Evans}}]{Guo2009}
\bibinfo{author}{\bibfnamefont{X.}~\bibnamefont{Guo}},
  \bibinfo{author}{\bibfnamefont{D.-J.} \bibnamefont{Liu}}, \bibnamefont{and}
  \bibinfo{author}{\bibfnamefont{J.~W.} \bibnamefont{Evans}},
  \bibinfo{journal}{J. Chem. Phys.} \textbf{\bibinfo{volume}{130}},
  \bibinfo{pages}{074106} (\bibinfo{year}{2009}).

\bibitem[{\citenamefont{da~Silva and de~Oliveira}(2012)}]{Silva2012}
\bibinfo{author}{\bibfnamefont{F.~E.} \bibnamefont{da~Silva}} \bibnamefont{and}
  \bibinfo{author}{\bibfnamefont{M.~J.} \bibnamefont{de~Oliveira}},
  \bibinfo{journal}{Comput. Phys. Commun.} \textbf{\bibinfo{volume}{183}},
  \bibinfo{pages}{2001} (\bibinfo{year}{2012}).

\bibitem[{\citenamefont{Varghese and Durrett}(2013)}]{Varghese2013}
\bibinfo{author}{\bibfnamefont{C.}~\bibnamefont{Varghese}} \bibnamefont{and}
  \bibinfo{author}{\bibfnamefont{R.}~\bibnamefont{Durrett}},
  \bibinfo{journal}{Phys. Rev. E} \textbf{\bibinfo{volume}{87}},
  \bibinfo{pages}{062819} (\bibinfo{year}{2013}).

\bibitem[{\citenamefont{Fiore and Landi}(2014)}]{Fiore2014}
\bibinfo{author}{\bibfnamefont{C.~E.} \bibnamefont{Fiore}} \bibnamefont{and}
  \bibinfo{author}{\bibfnamefont{G.~T.} \bibnamefont{Landi}},
  \bibinfo{journal}{\textit{Phys. Rev. E}} \textbf{\bibinfo{volume}{90}},
  \bibinfo{pages}{032123} (\bibinfo{year}{2014}).

\bibitem[{\citenamefont{Fiore}(2014)}]{Fiore2014a}
\bibinfo{author}{\bibfnamefont{C.~E.} \bibnamefont{Fiore}},
  \bibinfo{journal}{\textit{Phys. Rev. E}} \textbf{\bibinfo{volume}{89}},
  \bibinfo{pages}{022104} (\bibinfo{year}{2014}).

\bibitem[{\citenamefont{Pastor-Satorras
  et~al.}(2015)\citenamefont{Pastor-Satorras, Castellano, Van~Mieghem, and
  Vespignani}}]{Pastor-Satorras2015}
\bibinfo{author}{\bibfnamefont{R.}~\bibnamefont{Pastor-Satorras}},
  \bibinfo{author}{\bibfnamefont{C.}~\bibnamefont{Castellano}},
  \bibinfo{author}{\bibfnamefont{P.}~\bibnamefont{Van~Mieghem}},
  \bibnamefont{and}
  \bibinfo{author}{\bibfnamefont{A.}~\bibnamefont{Vespignani}},
  \bibinfo{journal}{Rev. Mod. Phys.} \textbf{\bibinfo{volume}{87}},
  \bibinfo{pages}{925} (\bibinfo{year}{2015}).

\bibitem[{\citenamefont{Ziff et~al.}(1986)\citenamefont{Ziff, Gulari, and
  Barshad}}]{Ziff1986}
\bibinfo{author}{\bibfnamefont{R.~M.} \bibnamefont{Ziff}},
  \bibinfo{author}{\bibfnamefont{E.}~\bibnamefont{Gulari}}, \bibnamefont{and}
  \bibinfo{author}{\bibfnamefont{Y.}~\bibnamefont{Barshad}},
  \bibinfo{journal}{\textit{Phys. Rev. Lett.}} \textbf{\bibinfo{volume}{56}},
  \bibinfo{pages}{2553} (\bibinfo{year}{1986}).

\bibitem[{\citenamefont{de~Oliveria et~al.}(2015)\citenamefont{de~Oliveria,
  da~Luz, and Fiore}}]{fiorePRE2015}
\bibinfo{author}{\bibfnamefont{M.~M.} \bibnamefont{de~Oliveria}},
  \bibinfo{author}{\bibfnamefont{M.~G.~E.} \bibnamefont{da~Luz}},
  \bibnamefont{and} \bibinfo{author}{\bibfnamefont{C.~E.} \bibnamefont{Fiore}},
  \bibinfo{journal}{Phys. Rev. E} \textbf{\bibinfo{volume}{92}},
  \bibinfo{pages}{062126} (\bibinfo{year}{2015}).

\bibitem[{\citenamefont{de~Oliveira et~al.}(2016)\citenamefont{de~Oliveira,
  M., Alves, and Ferreira}}]{Oliveira2016a}
  \bibinfo{author}{\bibfnamefont{M.~M.}~\bibnamefont{de~Oliveira}},
  \bibinfo{author}{\bibfnamefont{S.~G.} \bibnamefont{Alves}}, \bibnamefont{and}
  \bibinfo{author}{\bibfnamefont{S.~C.} \bibnamefont{Ferreira}},
  \bibinfo{journal}{\textit{Phys. Rev. E}} \textbf{\bibinfo{volume}{93}},
  \bibinfo{pages}{012110} (\bibinfo{year}{2016}).

\bibitem[{\citenamefont{Dias et~al.}(2014)\citenamefont{Dias, Ara\'ujo, and
  Telo da Gama}}]{diasPRE2014}
\bibinfo{author}{\bibfnamefont{C.~S.} \bibnamefont{Dias}},
  \bibinfo{author}{\bibfnamefont{N.~A.~M.} \bibnamefont{Ara\'ujo}},
  \bibnamefont{and} \bibinfo{author}{\bibfnamefont{M.~M.} \bibnamefont{Teloda
  Gama}}, \bibinfo{journal}{Phys. Rev. E} \textbf{\bibinfo{volume}{90}},
  \bibinfo{pages}{032302} (\bibinfo{year}{2014}).

\bibitem[{\citenamefont{Ara{\'u}jo et~al.}(2015)\citenamefont{Ara{\'u}jo, Dias,
  and Telo~da Gama}}]{araujo2015kinetic}
\bibinfo{author}{\bibfnamefont{N.~A.~M.} \bibnamefont{Ara{\'u}jo}},
  \bibinfo{author}{\bibfnamefont{C.~S.} \bibnamefont{Dias}}, \bibnamefont{and}
  \bibinfo{author}{\bibfnamefont{M.~M.} \bibnamefont{Telo~da Gama}},
  \bibinfo{journal}{J. Phys. Condens. Matter} \textbf{\bibinfo{volume}{27}},
  \bibinfo{pages}{194123} (\bibinfo{year}{2015}).

\bibitem[{\citenamefont{de~Oliveira et~al.}(2012)\citenamefont{de~Oliveira, Dos
  Santos, and Dickman}}]{Oliveira2012}
\bibinfo{author}{\bibfnamefont{M.~M.} \bibnamefont{de~Oliveira}},
  \bibinfo{author}{\bibfnamefont{R.~V.} \bibnamefont{Dos Santos}},
  \bibnamefont{and} \bibinfo{author}{\bibfnamefont{R.}~\bibnamefont{Dickman}},
  \bibinfo{journal}{\textit{Phys. Rev. E}} \textbf{\bibinfo{volume}{86}},
  \bibinfo{pages}{011121} (\bibinfo{year}{2012}).

\bibitem[{\citenamefont{Harris}(1974)}]{Harris1974}
\bibinfo{author}{\bibfnamefont{T.~E.} \bibnamefont{Harris}},
  \bibinfo{journal}{Ann. Probab.} \textbf{\bibinfo{volume}{2}},
  \bibinfo{pages}{969} (\bibinfo{year}{1974}).

\bibitem[{\citenamefont{de~Oliveira et~al.}(2014)\citenamefont{de~Oliveira,
   and Dickman}}]{Oliveira2014}
  \bibinfo{author}{\bibfnamefont{M.~M.}~\bibnamefont{de~Oliveira}},
  \bibinfo{author}{\bibfnamefont{R.}~\bibnamefont{Dickman}},
  \bibinfo{journal}{\textit{Phys. Rev. E}} \textbf{\bibinfo{volume}{90}},
  \bibinfo{pages}{032120} (\bibinfo{year}{2014}).

\bibitem[{\citenamefont{L\"{u}beck}(2003)}]{Luebeck2003a}
\bibinfo{author}{\bibfnamefont{S.}~\bibnamefont{L\"{u}beck}},
  \bibinfo{journal}{\textit{Phys. Rev. Lett.}} \textbf{\bibinfo{volume}{90}},
  \bibinfo{pages}{210601} (\bibinfo{year}{2003}).

\bibitem[{\citenamefont{Dickman et~al.}(1998)\citenamefont{Dickman, Vespignani,
  and Zapperi}}]{Dickman1998}
\bibinfo{author}{\bibfnamefont{R.}~\bibnamefont{Dickman}},
  \bibinfo{author}{\bibfnamefont{A.}~\bibnamefont{Vespignani}},
  \bibnamefont{and} \bibinfo{author}{\bibfnamefont{S.}~\bibnamefont{Zapperi}},
  \bibinfo{journal}{Phys. Rev. E} \textbf{\bibinfo{volume}{57}},
  \bibinfo{pages}{5095} (\bibinfo{year}{1998}).

\bibitem[{\citenamefont{Sander et~al.}(2016)\citenamefont{Sander, Costa, and
  Ferreira}}]{Sander2016}
\bibinfo{author}{\bibfnamefont{R.~S.} \bibnamefont{Sander}},
  \bibinfo{author}{\bibfnamefont{G.~S.} \bibnamefont{Costa}}, \bibnamefont{and}
  \bibinfo{author}{\bibfnamefont{S.~C.} \bibnamefont{Ferreira}},
  \bibinfo{journal}{Phys. Rev. E} \textbf{\bibinfo{volume}{94}},
  \bibinfo{pages}{042308} (\bibinfo{year}{2016}).

\bibitem[{\citenamefont{Assis and Copelli}(2009)}]{vladimirPRE2009}
\bibinfo{author}{\bibfnamefont{V.~R.~V.} \bibnamefont{Assis}} \bibnamefont{and}
  \bibinfo{author}{\bibfnamefont{M.}~\bibnamefont{Copelli}},
  \bibinfo{journal}{\textit{Phys. Rev. E}} \textbf{\bibinfo{volume}{80}},
  \bibinfo{pages}{061105} (\bibinfo{year}{2009}).

\end{thebibliography}
\end{document}